\documentclass[10pt]{wlscirep2}
\title{The value of ultrahigh resolution OCT in dermatology - delineating the dermo-epidermal junction, capillaries in the dermal papillae and vellus hairs}

\author[1,*]{Niels M{\o}ller Israelsen}
\author[1,2]{Michael Maria}
\author[3]{Mette Mogensen}
\author[3]{Sohpie Bojesen}
\author[1]{Mikkel Jensen}
\author[3]{Merete H\ae dersdal}
\author[2]{Adrian Podoleanu}
\author[1]{Ole Bang}
\affil[1]{Technical University of Denmark, DTU Fotonik, Kongens Lyngby, 2800, Denmark}
\affil[2]{University of Kent, School of Physical Sciences, Canterbury, Kent, England, CT2 7NZ}
\affil[3]{Department of Dermatology, Bisbebjerg Hospital, University of Copenhagen. Bispebjerg Bakke 23, DK-2400 Copenhagen NV, Denmark}

\affil[*]{nikr@fotonik.dtu.dk}

\begin{abstract}
Optical coherence tomography (OCT) imaging of the skin is gaining recognition and is increasingly applied to dermatological research.
A key dermatological parameter inferred from an OCT image is the epidermal (Ep) thickness as a thickened Ep can be an indicator of a skin disease. Agreement in the literature on the signal characters of Ep and the subjacent skin layer, the dermis (D), is evident. Ambiguities of the OCT signal interpretation in the literature is however seen for the transition region between the Ep and D, which from histology is known as the dermo-epidermal junction (DEJ); a distinct junction comprised by the lower surface of a single cell layer in epidermis (the stratum basale) connected to an even thinner membrane (the basement membrane). The basement membrane is attached to the underlying dermis. 

In this work we investigate the impact of an improved axial and lateral resolution on the applicability of OCT for imaging of the skin. To this goal, OCT images are compared produced by a commercial OCT system (Vivosight from Michaelson Diagnostics) and by an in-house built ultrahigh resolution (UHR-) OCT system for dermatology. 

In 11 healthy volunteers, we investigate the DEJ signal characteristics. We perform a detailed analysis of the dark (low) signal band clearly seen for UHR-OCT in the DEJ region where we, by using a transition function, find the signal transition of axial sub-resolution character, which can be directly attributed to the exact location of DEJ, both in normal (thin/hairy) and glabrous (thick) skin. To our knowledge no detailed delineating of the DEJ in the UHR-OCT image has previously been reported, despite many publications within this field. 

For selected healthy volunteers, we investigate the dermal papillae and the vellus hairs and identify distinct features that only UHR-OCT can resolve. Differences are seen in tracing hairs of diameter below $20\,\mu m$, and in imaging the dermal papillae where, when utilising the UHR-OCT, capillary structures are identified in the hand palm, not previously reported in OCT studies and specifically for \textit{glabrous} skin not reported in any other \textit{in vivo} optical imaging studies. 
\end{abstract}
\begin{document}

\flushbottom
\maketitle
%
%
\thispagestyle{empty}

\section{Introduction}
With two and a half decades of research and clinical work, optical coherence tomography (OCT) is a well-established medical tool used in diagnosis and understanding of numerous pathologies. In particular, OCT proved highly successful in diagnosing both poles of the eye \cite{huang1991optical,OCT2015}.

Less established is OCT in the field of Dermatology where scattering of light strongly limits the penetration into the skin \cite{welzel1997optical,avanaki2013quantitative}. Nevertheless, a few commercial systems have emerged and pushed \textit{in-vivo} studies in the clinics providing understanding of a variety of skin diseases where non-melanoma cancers, also termed keratinocyte carcinomas, are delineated and diagnosed by OCT with moderate to high diagnostic accuracy \cite{hussain2015optical,schuh2016optical}.

As the penetration depth is an important parameter in the attempt of fully identifying the extension of skin cancer, minimal signal loss in the primary skin layers from the surface is vital and therefore longer wavelengths (typically around 1300 nm) are chosen for skin imaging \cite{izatt1994optical,fujimoto1995optical,chin2016master}.

A second equally important parameter, pre-determining the information content of an OCT image, is the optical resolution, which presents a lower limit on the spatial details of the structures recognisable in the skin. Studies have shown that improving the resolution beyond regular OCT (lateral resolution 10-15 $\mu m$, axial resolution 5-10 $\mu m$) does allow one to extract an increasing number of pathological features otherwise only captured \textit{in vivo} by reflectance confocal microscopy (RCM) \cite{boone2014high,gambichler2015high}.  However, due to the heterodyning effect, of multiplication of the weak signal from skin with that of large power from the reference arm, the OCT penetration depth exceeds that of RCM. To improve the OCT even more in comparison with the RCM, it would be desirable to bring the OCT resolution to levels required for competing with histology \cite{mogensen2009assessment}. 

A number of ultra-high resolution (UHR)-OCT studies on malignant melanoma and non-melanoma skin lesions \textit{in vivo} have been performed with the OCT system \textit{Skintell} from AFGA providing a axial/lateral resolution in skin of $3\,\mu m$/$3\,\mu m$\cite{boone2014high,boone2015high1,boone2015high2,gambichler2015high,gambichler2015multicentre,boone2016vivo1,boone2016vivo2}. However only few studies address UHR-OCT visualising of \textit{healthy} skin \cite{boone2015high3,li2015epidermal} which may be as least as important in order to improve the non-pathological image characters, trends and benchmarks. Only a single study demonstrates a head to head clinical study comparing UHR-OCT with conventional OCT; this study concerns the OCT systems Callisto from Thorlabs (axial/lateral resolution in skin of $7\,\mu m$/$8\,\mu m$), Vivosight from Michelson Diagnostics (axial/lateral resolution in skin of $5\,\mu m$/$7.5\,\mu m$) and the ultrahigh resolution system Skintell introduced above \cite{schuh2016comparison}. This study however does not focus on the specific gain brought by the higher resolution when visualising skin layers of healthy skin but rather on the statistics generated in reading sizes of various skin structures or lesions detected. From these investigations they register decreased signal intensity and skin layer thicknesses in images generated by Skintell compared to Vivosight Callisto which they report to provide similar image details.

\subsection{The skin}
The outer layer of the skin, Epidermis (Ep), consists mainly of cells termed keratinocytes, tightly built together like bricks in a wall. The lower keratinocyte layer is called the basal cell layer and it is situated on top of the basement membrane, which separates the Ep from the Dermis (D) \cite{Weedon2002Skin}. An important parameter is the Ep thickness, where an abnormally thick Ep thickness can imply skin diseases such as psoriasis, eczema, and pre-cancerous dysplastic changes \cite{Weedon2002Skin}. The Ep thickness is a key benchmark and it is hence essential to determine the boundary between the Ep and the underlying D. The border zone termed the dermo-epidermal junction (DEJ) and represented by a single cell layer, the stratum basale, is also a site for skin diseases, in a particular the epidermolysis bullosa disorder \cite{nousari1999pemphigus,fine2014inherited}. The DEJ is thus an essential component and important to delineate both in order to extract exact measures of the Ep thickness but also to recognise pathological patterns of DEJ itself.

Several dermatological OCT studies have led to consensus on the DEJ being present somewhere in the transition between the Ep signal and the brighter D signal \cite{welzel1997optical,neerken2004characterization,weissman2004optical,gambichler2005epidermal}. The disordered nature of the DEJ was assumed to be unresolvable and contributing to the signal transition region as a laterally averaged phenomenon. In 2006, the first signal drop between the Ep (Ep) and dermis (D) signals was reported and subsequent studies started to some degree displaying and commenting on a signal minimum between the Ep and D signals \cite{gambichler2006vivo,mogensen2008morphology,blatter2012situ}. In 2012 the first dermatological studies applying UHR-OCT was published where a rich image labelling of the skin layers was presented \cite{blatter2012situ,lee2012three}. Clear visualisation of a signal drop between Ep and D was depicted in the UHR-OCT images but no detailed delineation of where exactly to draw the DEJ on the OCT image was presented in these or in the ensuing studies. In parallel, other studies presented the disagreement on where exactly to draw the DEJ within the Ep-D signal transition region \cite{hussain2015optical,kuo2016quantitative}. Several other groups did not introduce or explain their choice of DEJ definition in the OCT signal before
deducting the Ep thickness \cite{schuh2016optical,schuh2016comparison,taghavikhalilbad2017semi,adabi2017universal}.
\subsection{The study}
In this work we assess the clinical value of UHR-OCT in terms of delineation of the DEJ, small hairs and papillae of the D in a study involving 11 healthy voluntary participants (HPs). The images of a conventional commercially available multi-beam OCT system (C-OCT) and an in-house built (single focus) UHR-OCT system based on a supercontinuum source are compared side-to-side clinically.

First we delineate the signal drop between Ep and D signals, which we recognise as a low signal 'band' (abbreviated \textit{DB}) rather than a signal minimum. In doing so, we find an exact location of DEJ in the OCT image to the resolution limit of our system. Secondly, we trace two important types of adnexal structures, the vellus hairs in hair follicles and the papillae of the upper D in the palm. These are used comparatively by the two OCT systems to report on the  resolution difference.
%
%

\section{Methods}
\begin{figure}[ht!]
\centering\includegraphics[width=0.8\textwidth]{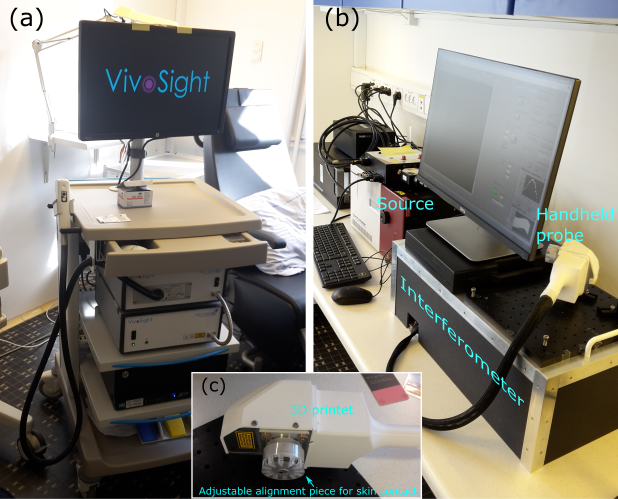}
\caption{Photographs of the C-OCT (a) and the UHR-OCT system (b) applied for the comparative study. (c) depicts the home-built handheld probe of the UHR-OCT system.}\label{Fig:systems}
\end{figure}

The two OCT systems used are a commercial (C-OCT) system, Vivosight, from Michelson, UK, and an in-house built system (UHR-OCT) which are seen in Fig.~\ref{Fig:systems}(a) and (b), respectively. 
\begin{table}
\centering
\begin{tabular}{|c|c|c|}
\hline 
\textbf{System feature} & \textbf{C-OCT} & \textbf{UHR-OCT} \\ 
\hline 
Operating wavelength (nm) & $1305$ & 1270  (1070-1470) \\ 
\hline 
Axial optical resolution in tissue ($\mu m$) & $<5$ & $2.2$ \\ 
\hline 
Lateral optical resolution in tissue ($\mu m$) & $<7.5$ & $4.6$ \\ 
\hline 
Depth of focus (mm) & 1 & 0.05 (Rayleigh length) \\ 
\hline 
$^*$Axial digital resolution in tissue ($\mu m$) & 4.12 & 1.46 \\ 
\hline 
$^*$Lateral digital resolution ($\mu m$) & 4.41 & 2.93 \\ 
\hline 
$^\dagger$Scanning area (mm $\times$ mm) & $4\times 4$ & $3\times 3$ \\ 
\hline  
$^\ddagger$Optical average power applied (mW) & $5\,mW$ & $5\,mW$\\
\hline
\end{tabular} 
\caption{Key characteristics of the OCT systems. Information on the C-OCT system is gained from the official website: $\mathtt{vivosight.com/researcher}$. The skin tissue is assumed to have a refractive index of $n=1.35$. *: Based on the fast-axis lateral pixel-step in the exported images.$\dagger$: The scanning area chosen for this specific study.$\ddagger$: Measured with photo-diode S122C from Thorlabs.}\label{characteristics}
\end{table}
Key parameters expressing the performance of the systems are displayed in table~\ref{characteristics}. The commercial system is a multi-beam system providing four different focal positions with a swept  source. The four focii, are positioned so they effectively form a $1\,mm$ depth of focus.

The UHR-OCT system is powered by a 320~MHz supercontinuum source (SuperK Extreme, EXR-9) from NKT Photonics. The high repetition rate provides significant averaging to reduce the noise and makes the source suitable for OCT \cite{moller2012power,maria2017q}. The source is bandpass filtered (1000-1500~nm) and light from the this is coupled into one arm of a regular Michelson interferometer and guided evenly to a reference arm and to a sample arm, the latter terminating with a handheld probe. The probe is home-built and customised for skin contact, see Fig.~\ref{Fig:systems}(c). The combined signal is detected by a spectrometer from Wasatch Photonics, USA, covering the spectrum 1074-1478~nm. The optical resolutions were inferred by imaging a USAF 1951 target and a plane mirror. The sensitivity is 89~dB (4~mW sample power). All images presented were build from a 76~kHz spectral line rate and, if not specified otherwise, no image averaging was performed.

The research protocol was approved by the Ethics Committee The Capital Region of Denmark: no. H-16039077. It was carried out at Bisbebjerg Hospital, Denmark, and is described as follows. 11 healthy voluntary participants (HPs) were investigated. Information on age, Fitz-Patrick skin type and gender is given in the \textit{Appendix A}, table~\ref{HPtable}. All HPs were scanned on the hand palm (beneath index and middle finger) and on the cheek, first with the C-OCT system and then with the UHR-OCT system. In between interchanging OCT device, skin areas were marked to retain the selected scanning area. In this way 22 skin volumes of cheeks and palms were obtained for each OCT device.  

Before image analysis, images from the C-OCT system were saved as tiff files (Tagged Image File Format), with 8-bit image depth, i.e. 256 grey scale levels in between black and white. The raw data, obtained from the UHR-OCT measurements, was re-sampled and digitally dispersion compensated \cite{makita2008full}, and finally shadow compensated to account for signal attenuation in depth caused by scattering \cite{hojjatoleslami2012oct,zhang2014dual}. In order not to compromise the axial resolution, no spectral windowing and no image smoothing was performed. Shadow compensation was applied in order to increase the contrast between the various skin layers and simultaneously found to suppress artefacts normally toned down by applying spectral windowing. An example of the contrast gain in utilising shadow compensation is given in Fig.~\ref{ShadowCompare}, \textit{Appendix C}.

Matlab post-processing of images from both systems was performed to define boundaries between the Ep and the D. In doing this, the skin surface was applied as a reference. As the surface is easily recognised as the layer with the strongest reflected signal, a simple peak-finder algorithm convolving each A-scan with a 2-5 pixel kernel was applied for surface recognition.

To determine signal valley characteristics in the axial dimension, a least squares fitting of a double sigmoid function was performed, as detailed in the \textit{Appendix B}. This was used under the assumption that in the boundary region of Ep and D and with signal in-depth attenuation accounted for, each of the two skin layers presents a constant signal level (if speckle averaged) forming signal plateaus on either side of the signal valley \cite{welzel1997optical}. To simplify the fitting, we additionally assumed the extension of the two slopes, forming the valley between Ep and D signals, to be equal and thereby locked in the fitting process. From the fit we extract two key parameters; the  \textit{dark band (DB)}, defined as the axial distance between the two slopes forming the valley, and the \textit{edge sharpness (ES)}, defined as the axial extension of the valley side. Examples of the fitting are given in the \textit{Results} section. 

All images displayed are axially scaled as in tissue assumed to have a refractive index of $n=1.35$. 

Additionally,  for highly reflective surfaces, we have found an artefact about $16\,\mu m$ below the surface signal, replicating the surface signal, in the UHR-OCT images. We associate this with the subsurface sidelobe of the surface signal seen for a non-gaussian spectrum in the Fourier domain. We expect sidelobes in the image from the strongest signals as no windowing of the UHR-OCT spectra was performed as stated above, and specifically the subsurface sidelobe is enhanced relative to the above-surface sidelobe due to the shadow compensation algorithm applied.

\section{Results}
\begin{figure}[h!]
\centering\includegraphics[width=1.0\textwidth]{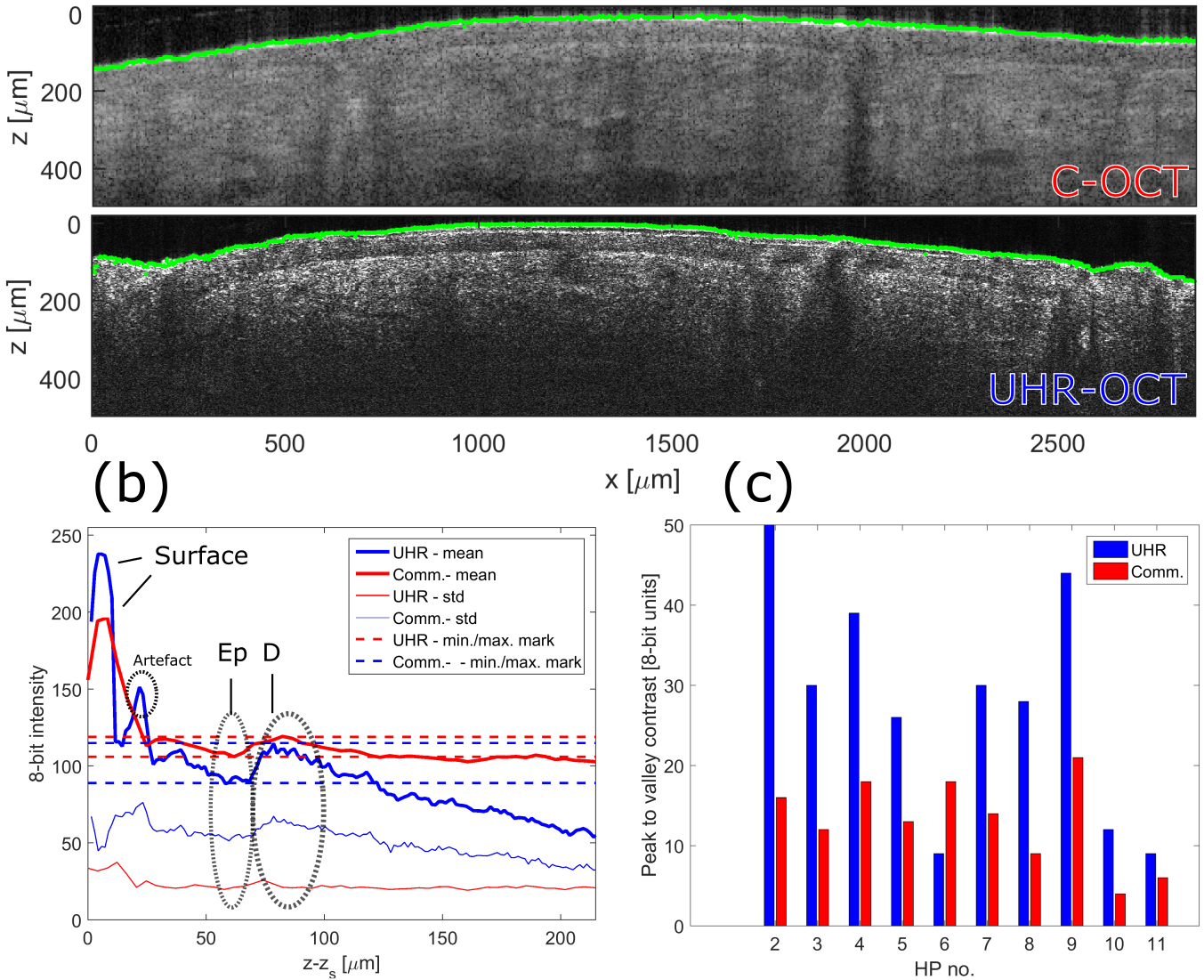}
\caption{(a): OCT images from HP 5 of the cheek generated with the C-OCT and the UHR-OCT system with the green delineations marking the surface tracing performed and representing the axial positions $z_S$. (b) Image signal average along 'x' relative to the surface trace. The horisontal dashed lines mark the signal readings of Ep and D utilised for computing the Ed-D contrast. (c) presents the Ep-D contrasts calculated for central B-scans each associated with a HP volume scan. HP 1 is excluded due to crucial artefacts in the scans.}\label{cheek1}
\end{figure}
\begin{figure}[h!]
\centering\includegraphics[width=1.0\textwidth]{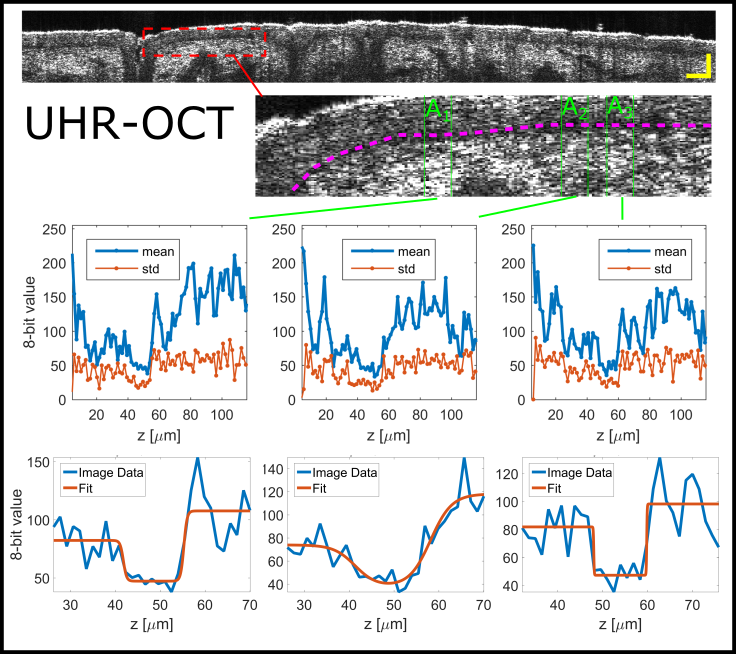}
\caption{UHR-OCT system: DEJ evaluation for central B-scan of the cheek of HP 2 (top image), scale bar representing 100~$\mu$m. A ROI is selected (dashed rectangle) of which a zoom-in is presented (bottom image). Within the zoom-in three subsets (A$_1$, A$_2$ and A$_3$) are emphasised (green vertical lines) and a guide to the eye of the DB propagation through the skin (purple dashed) is given. The subsets, comprising each, ten A-scans and representing regions of constant axial DEJ position, are averaged laterally (horizontally in image) to accentuate the DEJ from the speckle noise providing three graphs with means and stds (top graphs). A double sigmoid fitting (bottom graphs) is performed for each average to extract DEJ information.}\label{UHRcheek}
\end{figure}
\begin{figure}[h!]
\centering\includegraphics[width=1.0\textwidth]{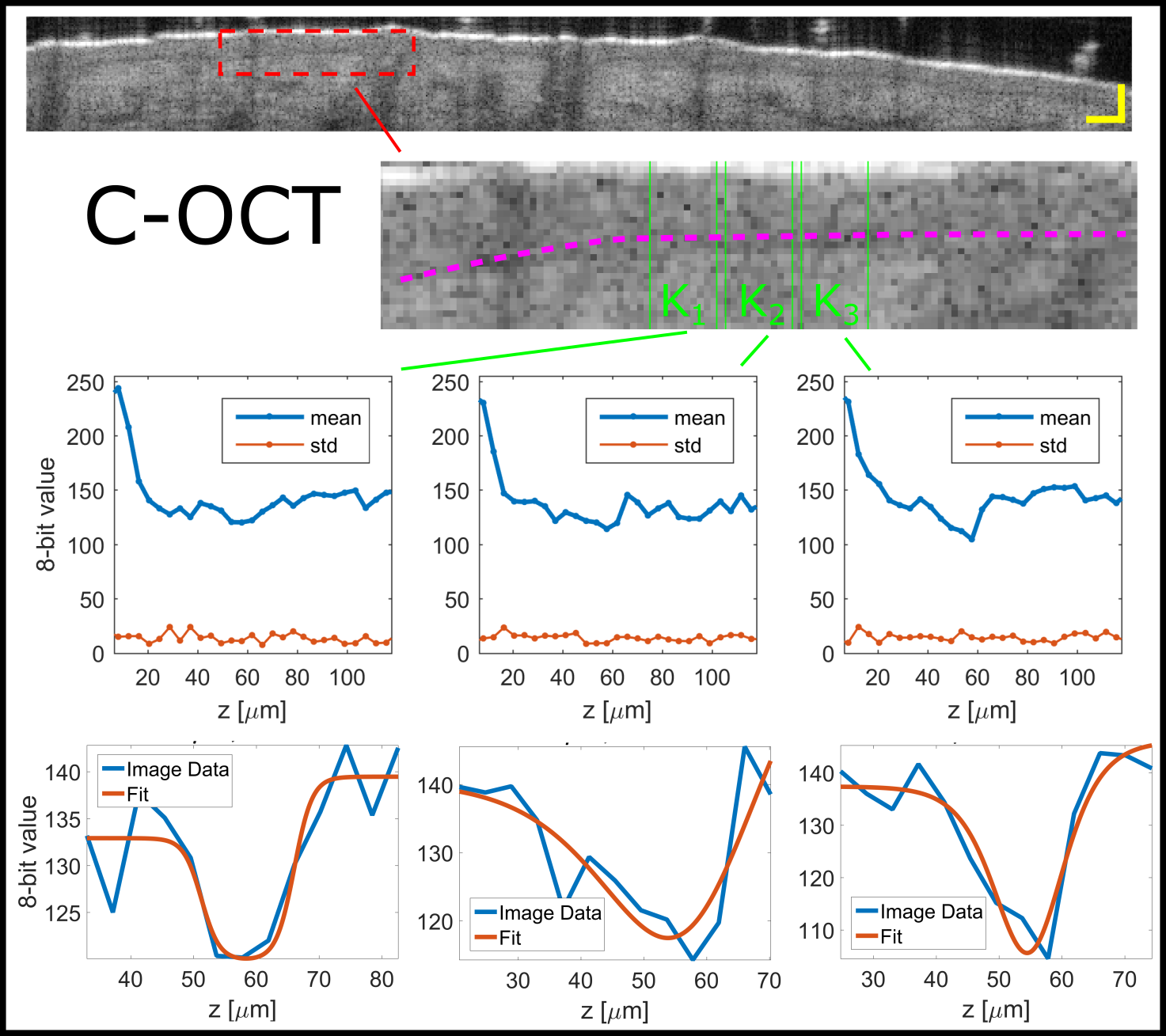}
\caption{C-OCT system: DEJ evaluation for central B-scan of the cheek of HP 2 (top image), scale bar representing 100~$\mu$m. A ROI is selected (dashed rectangle) of which a zoom-in is presented (bottom image). Within the zoom-in three subsets (K$_1$, K$_2$ and K$_3$) are emphasised (green vertical lines) and a guide to the eye of the DB propagation through the skin (purple dashed) is given. The subsets, comprising each, ten A-scans and representing regions of constant axial DEJ position, are averaged laterally (horizontally in image) to accentuate the DEJ from the speckle noise providing three graphs with means and stds (top graphs). A double sigmoid fitting (bottom graphs) is performed for each average to extract DEJ information.}\label{commcheek}
\end{figure}
\begin{figure}[ht!]
\centering\includegraphics[width=1.0\textwidth]{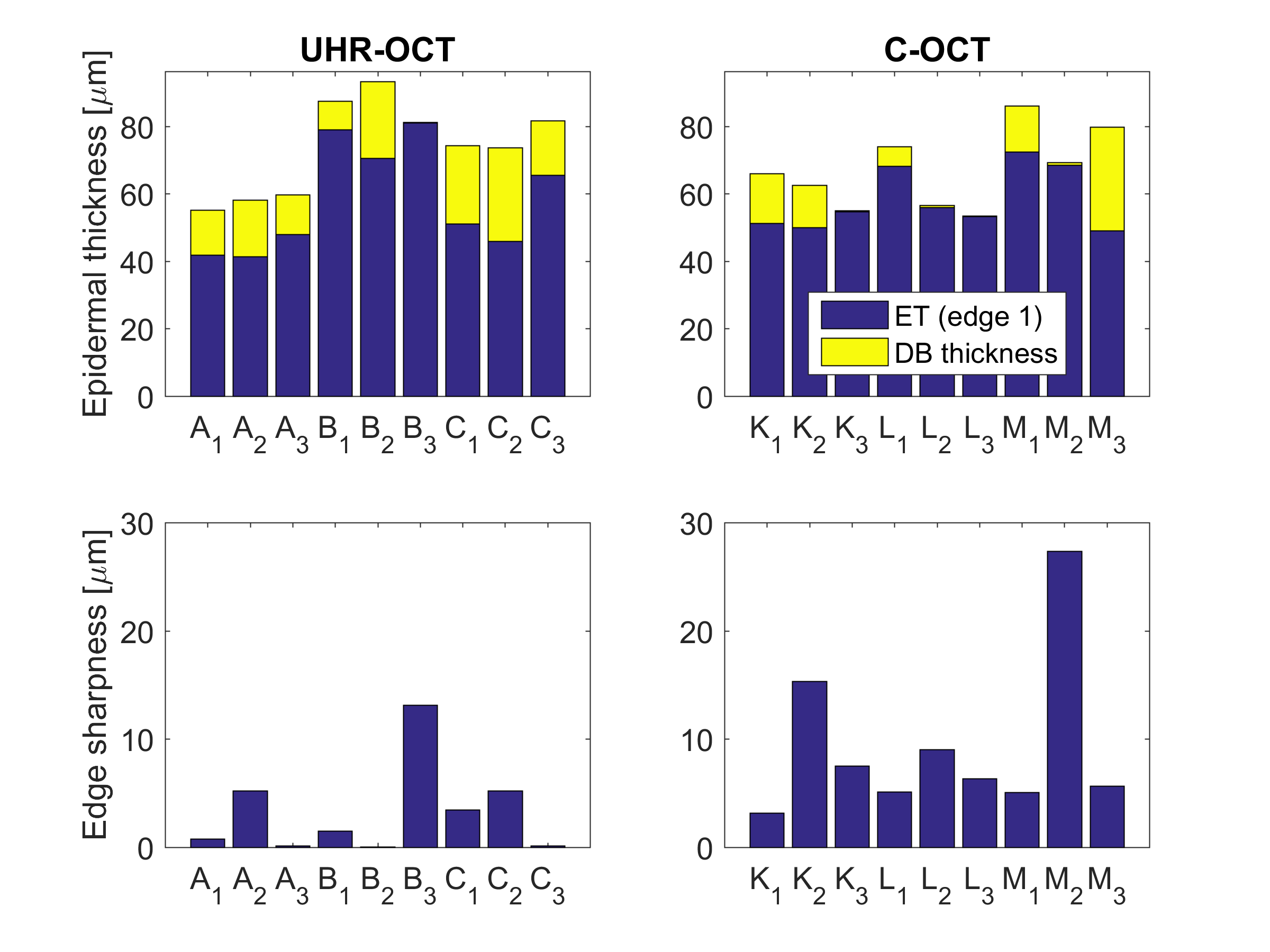}
\caption{The cheek: Ep thickness values and DEJ edge sharpness values found from the double sigmoid fitting procedure for both C-OCT and UHR-OCT system central B-scans for the cheek. A, B and C (K, L and M) denote the three subsets of each of the three high-contrast B-scans denoted 1, 2 and 3 (HP2, HP4 and HP9).}\label{CheekEval}
\end{figure}

\begin{figure}[ht!]
\centering\includegraphics[width=1.0\textwidth]{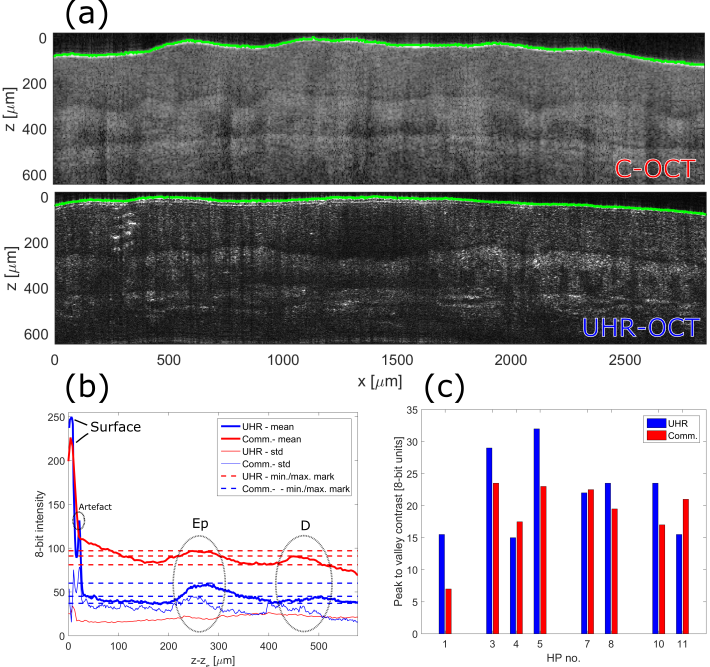}
\caption{(a): OCT images from HP 7 of the hand palm generated with the C-OCT and the UHR-OCT system with the green delineations marking the surface tracing performed. (b) Image signal average along 'x' relative to the surface trace. The horisontal dashed lines mark the signal readings of Ep, DEJ and D utilised for computing the Ed-D contrast. (c) presents the Ep-D contrasts calculated for central B-scans each associated with a HP volume scan. HP 2, HP 6 and HP 9 are excluded due to significant deviations in the DEJ relative position in the scans caused by special skin features.}\label{palm1}
\end{figure}
\begin{figure}[ht!]
\centering\includegraphics[width=1.0\textwidth]{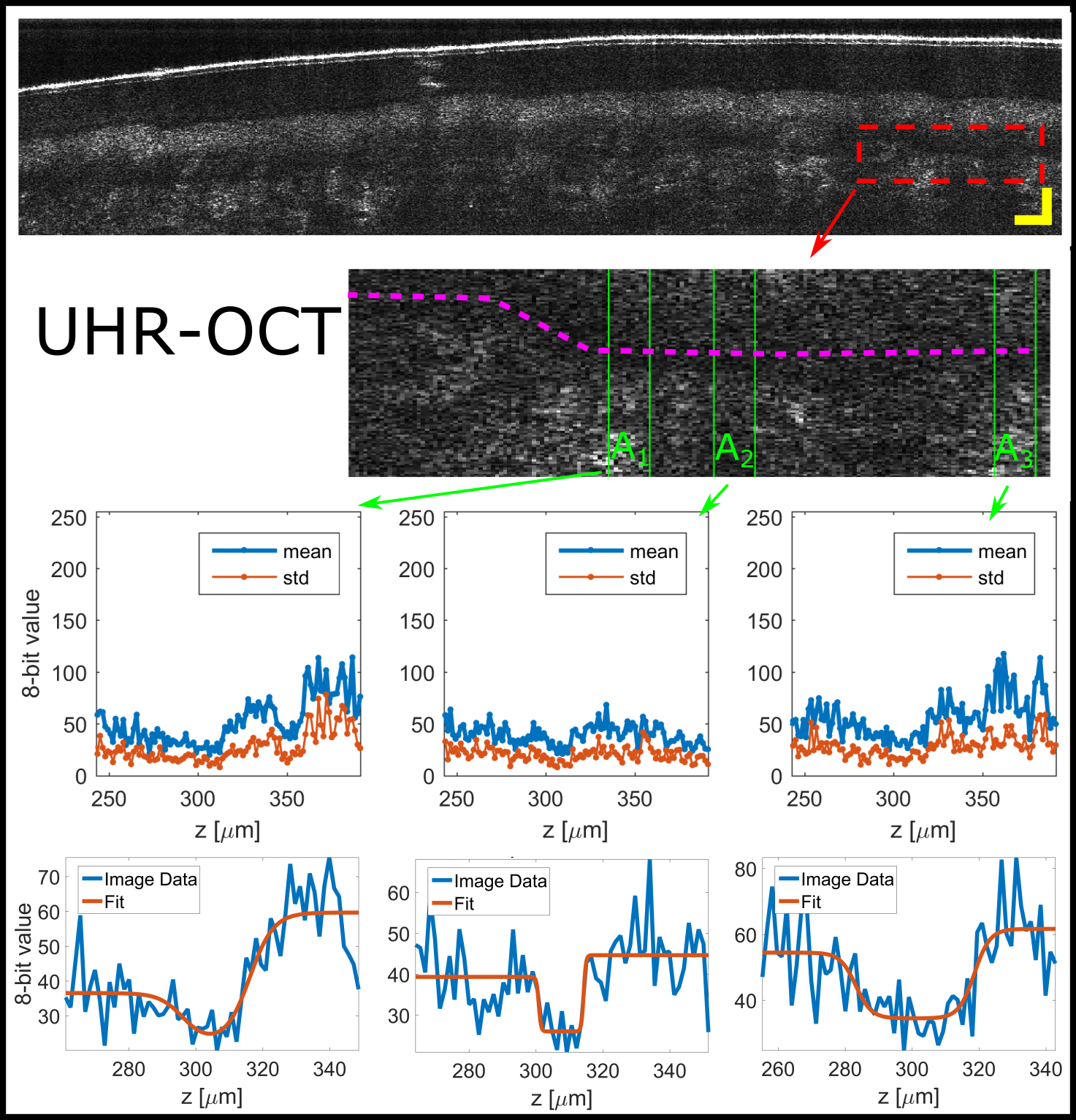}
\caption{UHR-OCT system: DEJ evaluation for central B-scan of the hand palm of HP 3 (top image), scale bar representing 100~$\mu$m. A ROI is selected (dashed rectangle) of which a zoom-in is presented (bottom image). Within the zoom-in three subsets (A$_1$, A$_2$ and A$_3$) are emphasised (green vertical lines) and a guide to the eye of the DB propagation through the skin (purple dashed) is given. The subsets, comprising each, ten A-scans and representing regions of constant axial DEJ position, are averaged laterally (horizontally in image) to accentuate the DEJ from the speckle noise providing three graphs with means and stds (top graphs). A double sigmoid fitting (bottom graphs) is performed for each average to extract DEJ information.}\label{UHRpalm}
\end{figure}
\begin{figure}[ht!]
\centering\includegraphics[width=1.0\textwidth]{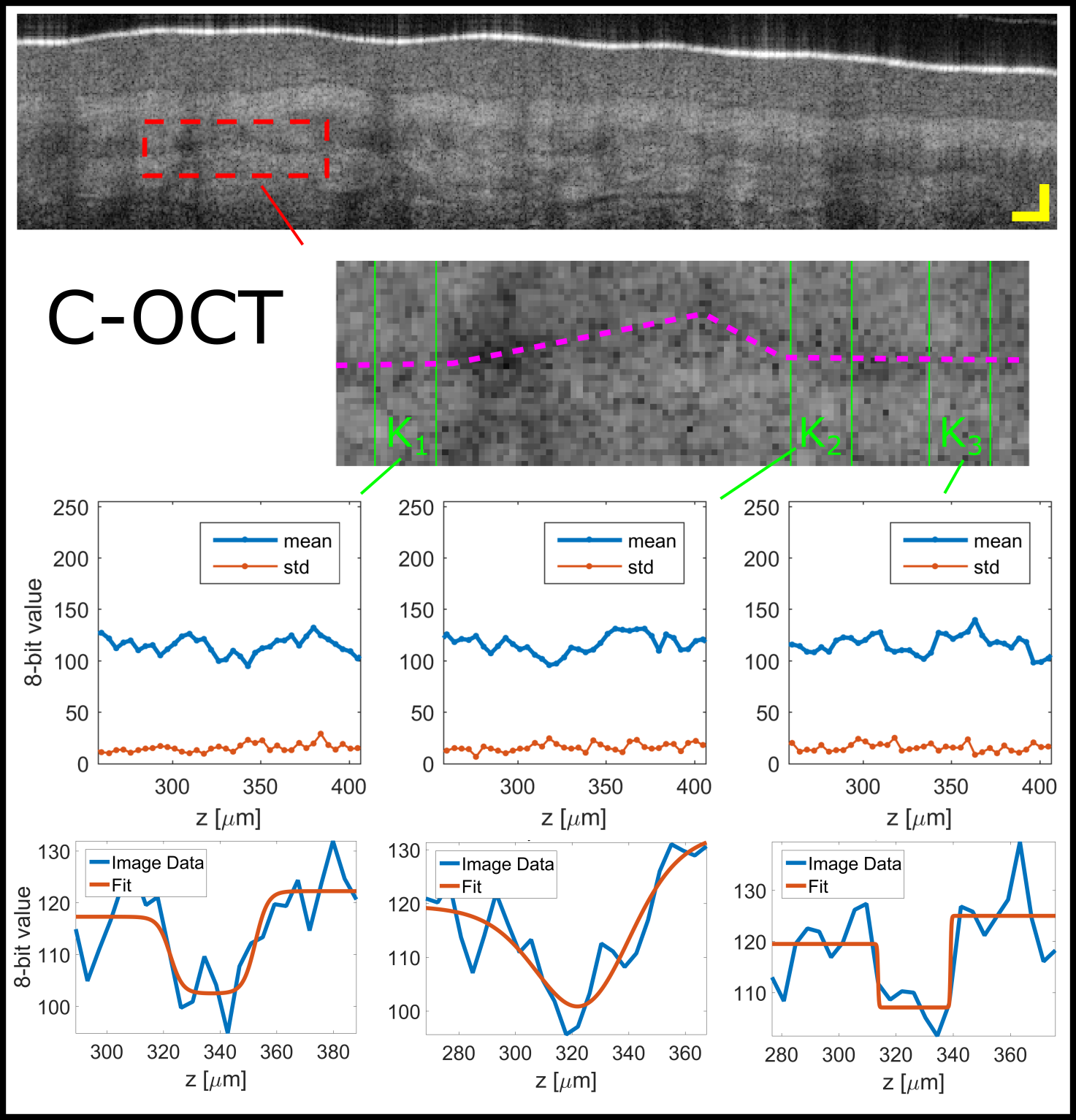}
\caption{C-OCT system: DEJ evaluation for central B-scan of the hand palm of HP 3 (top image), scale bar representing 100~$\mu$m. A ROI is selected (dashed rectangle) of which a zoom-in is presented (bottom image). Within the zoom-in three subsets (K$_1$, K$_2$ and K$_3$) are emphasised (green vertical lines) and a guide to the eye of the DB propagation through the skin (purple dashed) is given. The subsets, comprising each, ten A-scans and representing regions of constant axial DEJ position, are averaged laterally (horizontally in image) to accentuate the DEJ from the speckle noise providing three graphs with means and stds (top graphs). A double sigmoid fitting (bottom graphs) is performed for each average to extract DEJ information.}\label{commpalm}
\end{figure}

\begin{figure}[ht!]
\centering\includegraphics[width=1.0\textwidth]{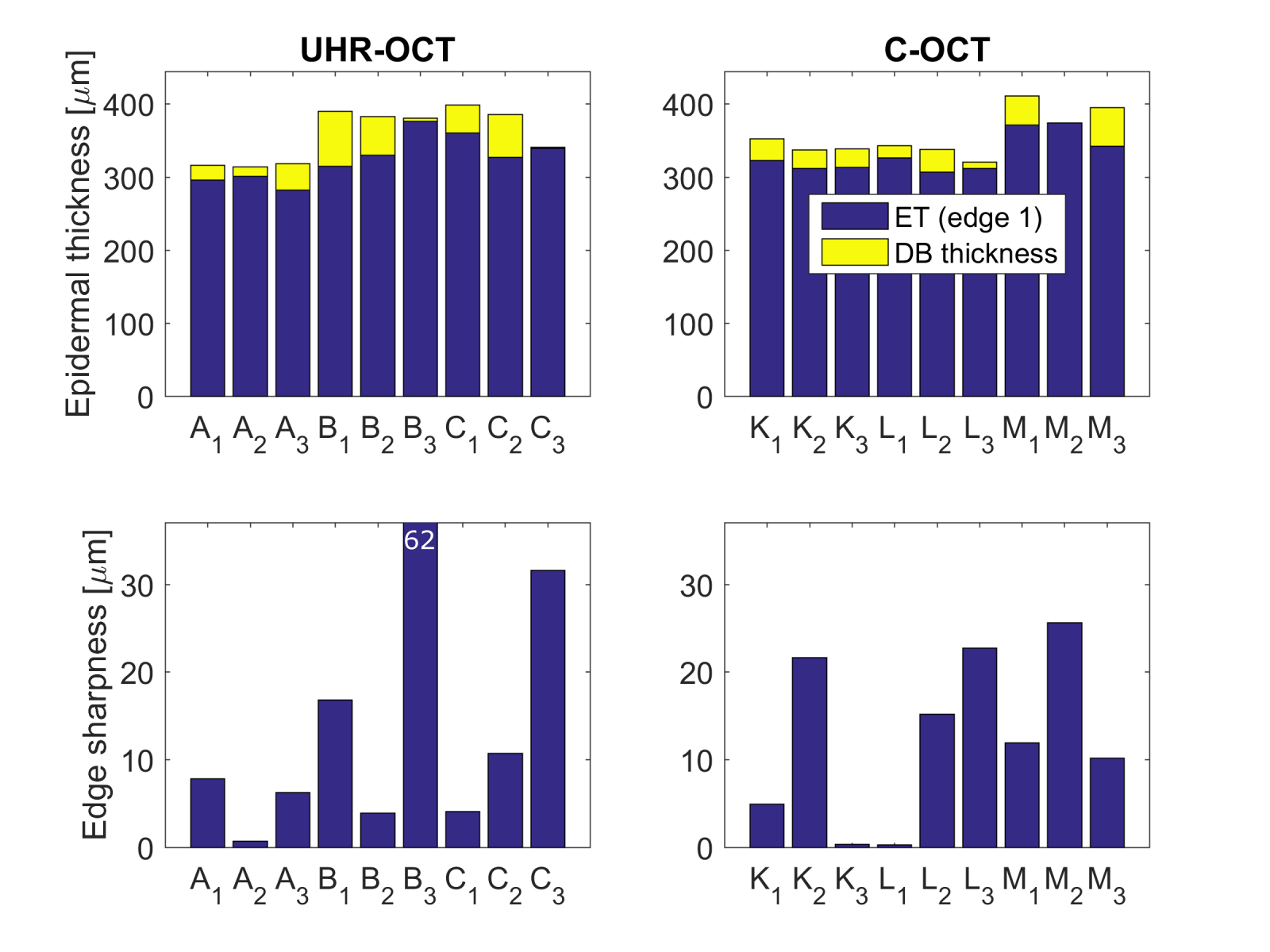}
\caption{The palm: Ep thickness values and DEJ edge sharpness values found from the double sigmoid fitting procedure for both C-OCT and UHR-OCT system central B-scans for the palm. A, B and C (K, L and M) denote the three subsets of each of the three high-contrast B-scans denoted 1, 2 and 3 (HP2, HP4 and HP9).}\label{palmEval}
\end{figure}

\begin{figure}[ht!]
\centering\includegraphics[width=.9\textwidth]{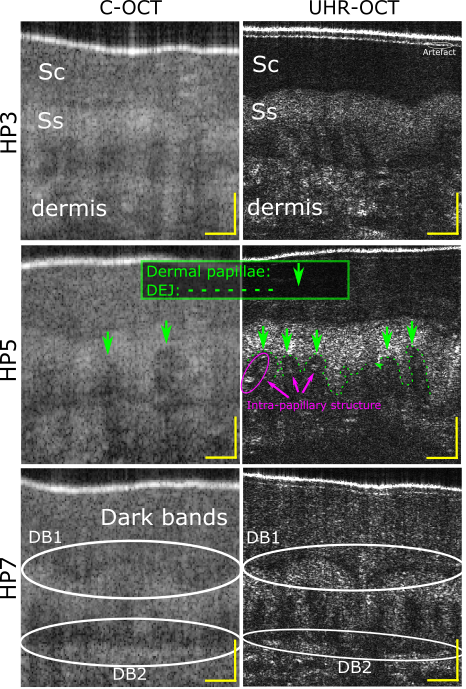}
\caption{B-scan zoom-in highlighting skin structures in the palm of the high contrast HPs for C-OCT and UHR-OCT. Stratum corneum (Sc) and stratum spinosum (Ss) and D is marked for HP3. Dermal papillaries and the exact DEJ is marked for HP5. The characteristic two dark bands (DB1 and DB2) seen for OCT signals of glabrous skin are surrounded by oval contours. The scale bars represent $100\mu m$.}\label{pap}
\end{figure}
\begin{figure}[ht!]
\centering\includegraphics[width=0.7\textwidth]{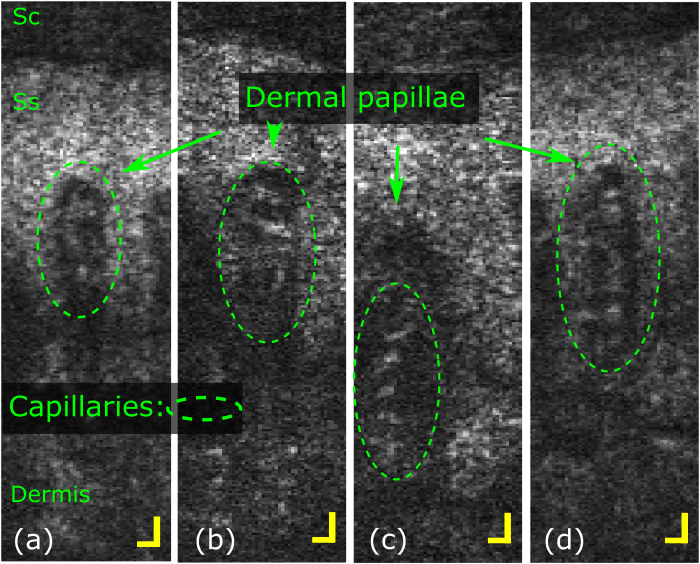}
\caption{Zoom-ins on individual dermal papillae of the palm of HP5. (a)-(d) depicts capillaries of the dermal papillae seen in the UHR-OCT images generated from 3-5 averaged B-scans each. Scale bars are $20\,\mu m$.}\label{Cap}
\end{figure}
\begin{figure}[ht!]
\centering\includegraphics[width=1.0\textwidth]{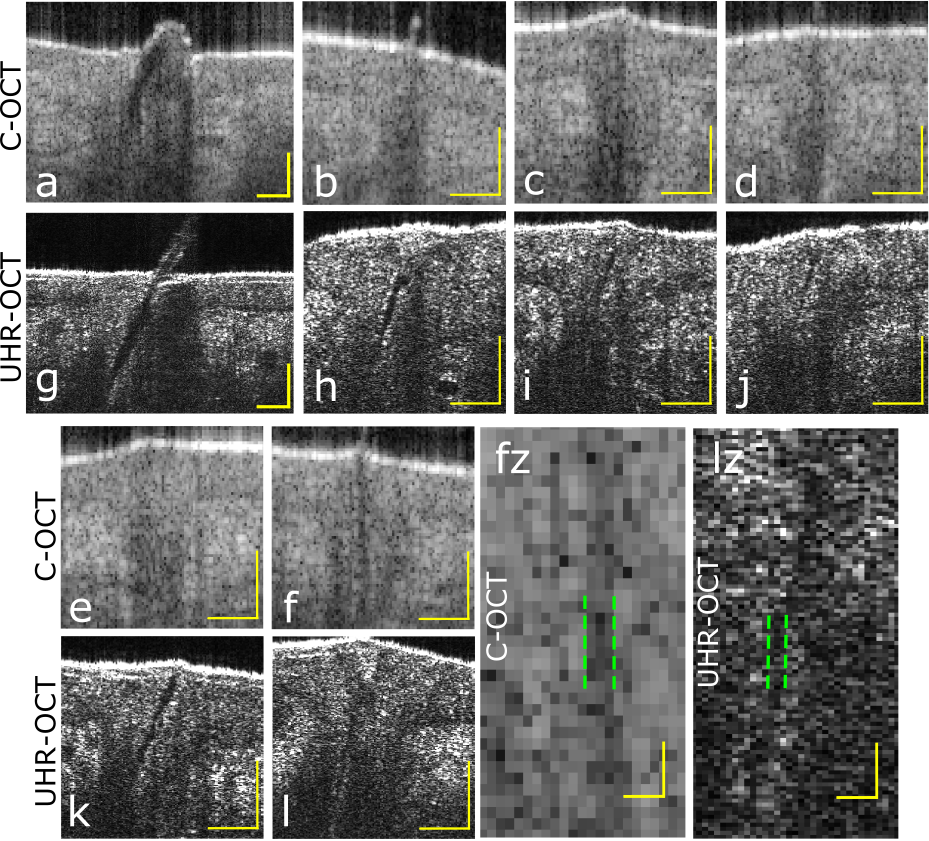}
\caption{Imaging of smaller and larger hairs on cheek of HP 4 comparing C-OCT and UHR-OCT systems. In images a-f and g-l hair follicles are located, each with scale bars representing 100~$\mu$m. a and g show larger hairs. fz and lz are zoom-ins of f and l, respectively, and pairs of vertical dashed green lines mark hair-to-surroundings bounderies, each pair enclosing three rows of pixels. Scale bars in fz and lz represent 20~$\mu$m.}\label{hairs}
\end{figure}	
\subsection{Imaging of the dermo-epidermal junction}
The first comparison was performed in the frame of recognising and evaluating the DEJ. For this evaluation one B-scan of each volume was selected as a representative image. The selection was made in the way that the central B-scan (along the slow scanning axis) of each skin volume was chosen. First a relative contrast of the selected B-scans was estimated. This was done as follows:

First alignment of the individual A-scans to each other, with the surface as reference, was achieved. Subsequently the new surface-levelled B-scan was averaged along the lateral dimension. The resultant 1-dimensional depth profile served as a view of the in-depth signal trends. This was performed under the assumption that 1: the Ep thickness is approximately constant with the disordered DEJ leading only to a smearing of D and Ep signal characters in a smaller axial region (compared to the Ep thickness) and 2: no adnexal structures takes up more than $\sim 10\%$ of the lateral B-scan range with the potential of distorting the Ep and D signals. If significant adnexal structures (i.e. hairs) were found in any selected B-scans, these cases were discarded.
Two examples (one from each OCT system) of selected B-scans of the cheek are seen in Fig.~\ref{cheek1}(a). 
The resultant lateral averaging is shown in Fig.~\ref{cheek1}(b) showing the mean and standard deviation (std). As an OCT depth reflectivity profile in normal skin is normally characterised as a hyper-reflective signal (the surface), a lower signal (epidermal region) and yet higher signal (dermal region) \cite{welzel1997optical,neerken2004characterization}, we define the first sub-surface valley (between $z-z_s=50\,\mu m$ and $z-z_s=75\,\mu m$ in Fig.~\ref{cheek1}(b)) as Ep and the first sub-surface peak (between $z-z_s=75\mu m$ and $z-z_s=100z\mu m$) as D. The difference in maximum signal (D) and minimum signal (Ep), marked in Fig.~\ref{cheek1}(b) with dashed lines, is defined as the contrast across the DEJ. In this fashion, a DEJ contrast for each HP and OCT system was derived. The results of this investigation on the cheek is presented in Fig.~\ref{cheek1}(c) as a histogram showing the peak-to-valley DEJ contrast in 8-bit units. A-scan averages like the one presented in Fig.~\ref{cheek1}(b) for all HPs can be found in the \textit{Appendix D}. We note that apart from HP6, where the focus position was poor, the UHR-OCT system provides about double the contrast to that achieved by the C-OCT system.

To compare the DEJ regions in more detail for the two systems, three HP cases were singled out. To not favour our in-house built system, the HP cases with the three highest contrast for the C-OCT system were chosen for further studies, namely HP2, HP4 and HP9. For the six cheek B-scans of these HPs, a region of interest (ROI) was selected which represented a lateral range with a clear Ep and D signal and with no adnexal structures complicating the recognition of the DEJ region. 
An example of this, presented for HP2, is given in Fig.~\ref{UHRcheek} and Fig.~\ref{commcheek} marked with a punctured red rectangle, respectively. Within each ROI, three lateral subsets were selected for UHR-OCT, A$_x$, B$_x$ and C$_x$ (HP2, HP4 and H9), and C-OCT,  K$_x$, L$_x$ and M$_x$ (HP2, HP4 and H9), with 'x' denoting the subset number (1 to 3). The subsets were selected as areas where the DEJ position in the axial dimension appeared identical. Each subset comprised ten A-scans, as marked for each subset with two green lines in Fig.~\ref{UHRcheek} and Fig.~\ref{commcheek} for the UHR-OCT and C-OCT, respectively. By laterally averaging these, a smoothened profile across the DEJ region was generated, also presented in Fig.~\ref{UHRcheek} and Fig.~\ref{commcheek}.   
%
Finally the dark band (DB) recognised between the Ep signal and D signal was evaluated by fitting a double sigmoid function to the profiles obtained from the subsets in order to estimate the width and the slope extension of the dark band valley in the DEJ region. Examples of the fittings are seen in Fig.~\ref{UHRcheek} and Fig.~\ref{commcheek}, bottom. 

From the three high-contrast HP cases selected, we infer nine measures of the DB separating the Ep and D signals for each OCT system. Numbers on the Ep thickness, the DB and the DB edge sharpness (ES), as defined in the \textit{Appendix B} (see Fig.~\ref{ModelCurve}), are summarised in Fig.~\ref{CheekEval}.
Comparing the UHR-OCT histograms (A, B and C, 1 to 3) we see that apart from B$_3$, exhibiting sub-micron DB and $13\,\mu m$ ES, the subsets agree on a 'dark band' region of widths between $8\,\mu m$ and $28\,\mu m$ in between the two higher signal regions corresponding to Ep and D. C-OCT however have four subsets ($K_3$, $L_2$, $L_3$, and $M_2$) detecting vanishing ($1\,\mu m$ and smaller) DBs, where the remainder five detect DBs in the range between $5\,\mu m$ and $31\,\mu m$. For the DEJ ESs, defined as the axial extension of the Ep/DB and DB/D signal transitions (detailed definition in \textit{Appendix B}), there is also a significant difference between the results of the systems.  All UHR-OCT subsets, apart from B$_3$, show $6\,\mu m$ or smaller ESs and for C-OCT the majority (all but, K$_1$) show ESs of $5\,\mu m$ or larger.

Likewise the measurements on the hand palm, representing glabrous skin, was analysed. Characteristic of glabrous skin is the modified and extraordinary thick Ep. The stratum corneum is the outermost layer of the Ep and consists of dead cells, corneocytes, that have shed interior organelles. On palms and soles the stratum corneum is more than 10 times thicker than elsewhere on the body \cite{Weedon2002Skin}. The main component is the protein keratin.

Examples of palm images produced by both systems are displayed in Fig.~\ref{palm1}(a). Here we see the extraordinary thick stratum corneum extending about $200\,\mu m$ into the skin. Unlike for the cheek, the DEJ region is sandwiched between parts of Ep and D providing equal signal strengths and is more vaguely recognised as a lower signal valley.      
With this structure we instead trace two signal peaks and the signal valley in between to construct a contrast of the DEJ region. An example of this approach is seen in Fig.~\ref{palm1}(b). The signal peak-valley-peak contrast evaluation summary for all HPs, without serious image artefacts, is given in Fig.~\ref{palm1}(c). Opposite to the cheek analysis, no clear tendency on which OCT system performs better in the frame of the peak/valley contrast metric is observed with UHR-OCT providing the best DEJ contrast in five out of eight HP cases. 

As for the cheek, the HP cases with the highest C-OCT contrast were more closely inspected where ROIs were found for each HP image of each OCT system and consecutively three subsets representing constant axial position of the DEJ region were selected. Examples for both the UHR-OCT and the C-OCT are depicted in Fig.~\ref{UHRpalm} and Fig.~\ref{commpalm}.  

The complete evaluation of the nine double sigmoid fittings to the DEJ regions of the selected image subsets is presented in Fig.~\ref{palmEval}. For the hand palm both UHR-OCT and C-OCT system detects from
vanishing thin DBs and up to $75\,\mu m$ and $53\,\mu m$ thick DBs, respectively. The spread on ESs is also large, both within a single HP case and generally for both systems, ranging from a few microns to tens of microns. The large fluctuations in Ep thickness and ES between subsets for the C-OCT is considered due to the lack in resolution and contrast. For the UHR-OCT the lack of contrast is the critical parameter which is evident when comparing Fig~\ref{UHRcheek} and Fig~\ref{UHRpalm} and the resultant ESs in Fig~\ref{CheekEval} and Fig~\ref{palmEval}, for which the DEJ region spans a much larger range of 8-bit units.

\subsection{Imaging of dermal papillae}
A characteristic skin structure associated with DEJ is the papillae in the D. The DEJ rests on top of the dermal papillae containing the characteristic capillary loops that supply the Ep with oxygen. The dermal papillae are finger-like protrusin (papillae) from D into Ep, or undulations between Ep and D. In some parts of the skin they are more flattened. The degree of flattening also depends on age where the number of papillae are reduced with increasing age \cite{mizukoshi2015changes,newton2015skin}. 

Images of the papillae are given for the two systems in Fig.~\ref{pap} detected in the palm volumes of the high-contrast HPs defined above (HP3, HP5 and HP7). The figure displays papillae for the three HPs where the resolution, both axially and laterally is probed. In general the papillae seen for C-OCT appears more blurry and with poorer contrast compared to UHR-OCT, where also intra-papillary structures can be recognised (HP5). 

When averaging three to five UHR-OCT B-scans (corresponding to $9\,\mu m$ to $15\,\mu m$, laterally), we can infer the intra-papillary  structure as presented in Fig.~\ref{Cap}. Here, we see blood vessel-like structures rising from the dermis and terminating in a curling fashion, each within a host papillary. One structure shows corkscrew like body (Fig.~\ref{Cap}(c)) where the remaining three show more randomly curled structures (Fig.~\ref{Cap}, (a), (b) and (d)),

%
%
%
%
Also marked in Fig.~\ref{pap} is the previously evaluated DB of the DEJ region (DB2 in Fig~\ref{pap}, HP7) as well as a \textit{second} DB, DB1 in Fig~\ref{pap}, HP7, found between the stratum corneum and stratum spinosum. For DB1 the UHR-OCT reveals that also the signal dip between these two layers is indeed a DB with 'sharp' edges.
\subsection{Imaging of the vellus hairs of the cheek}
To test the efficacy of the resolution on a second 2-dimensional structure, also relevant in context of diagnosing and delineating skin cancer, comparisons on resolving cheek hairs was performed. The subject of comparison was HP4 containing many small hairs (more than 50 hairs in each volume). Fig.~\ref{hairs} shows a number of detected hair follicles. The hair follicles are recognised in drops of Ep reaching beyond the penetration depth. Within the hair follicles hairs can be recognised in both C-OCT and UHR-OCT images as seen in the image example presented in Fig.~\ref{hairs}(a) and Fig.~\ref{hairs}(g), displaying relatively large hairs (widths $40-50\mu m$). The hairs are delineated as dark structures penetrating the surface in a skewed manner where almost no scattering is seen from. However, in most cases the hairs cannot be detected inside the hair follicles for the C-OCT (Fig.~\ref{hairs}(b-e)). The smallest hair tracked for C-OCT is depicted in Fig.~\ref{hairs},(f), where the projection and B-scan cut is just so, that the hair appears to penetrate orthogonally to the surface. For the UHR-OCT on the other hand, the small hairs can just be recognised as slim dark traces in almost all hair follicles observed with the characteristic tilting manner in penetration of the skin. These are seen in Fig.~\ref{hairs},(g)-(l) having diameters of $20\,\mu m$ or less. 

The smallest hairs of HP4, detected by the two systems, are the ones depicted in Fig.~\ref{hairs}(f) and Fig.~\ref{hairs}(l), where zoom-ins of these are provided in (fz) and (lz). For both C-OCT and UHR-OCT. the smallest hairs encompasses three pixel columns corresponding to widths of $13\,\mu m$ and $9\,\mu m$, respectively.


\section{Discussion}

%
\subsection{Healthy normal and glabrous skin}
Histology is to date the golden standard in terms of describing skin microstructure, such as cells and connective tissue in skin biopsies \textit{ex vivo}.

\begin{figure}[h!]
\centering\includegraphics[width=0.9\textwidth]{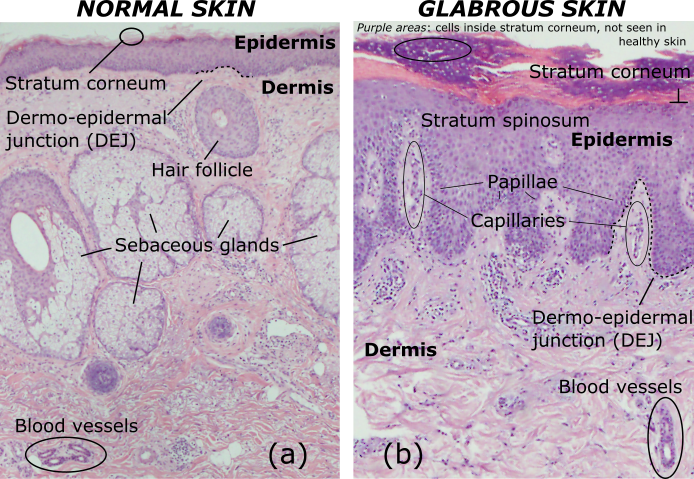}
\caption{Histology images of human skin representing the golden standard of histology. (a) is a histology projection of normal skin of the cheek and (b) presents a glabrous skin histology image of the palm with associated adnexal structures including dermal papillae capillaries. Images by courtesy of R. H. Nielsen, Rigshospitalet, Denmark.}\label{tyndtykhud}
\end{figure}	
Despite that the technique has evolved and improved through the years, it is however still an \textit{ex vivo} approach, and only represents the \textit{in vivo} state of the skin, with deformation of the skin associated with extracting a biopsy. Biopsies can be analysed from frozen sections or from dyed histology sections. Both methods result in some derformation compared to intact skin, usually shrinkage-related \cite{sandby2003epidermal}. It is therefore important to translate the rich information on anatomy from histology visualisation to the coarser resolved, however \textit{in vivo} visualisation provided by an OCT system. 

The general skin structure accepted on the basis of histology is extensively described in textbooks \cite{Elder2015Lever}. An example of normal and glabrous skin histology images is presented in Fig.~\ref{tyndtykhud}, where details at the cell-level stand out. 

The figure represents the skin layer components one ought to recognise in an OCT image in order to supplement the information obtainable by histology. Most prominent are Ep and D and the DEJ separating these. Additionally, hair follicles and sebaceous glands are visible for the normal skin. The glabrous skin is recognised by the thick stratum corneum, the irregular DEJ attributed to the strongly serrated dermal papillae, where also capillaries of these are visible here.    

\subsection{The dermo-epidermal junction in OCT}
As sketched in Fig.~\ref{tyndtykhud} and already described, the two skin regions, Ep and D are separated by the DEJ.
In OCT, the approximate relative axial position of DEJ has been known since 1997, where Welzel et al. launched the first dermatological OCT study \cite{welzel1997optical}. At the time they reported on the increase in signal going from the supposed Ep to the region believed to be D. On this basis they concluded that  the DEJ axial position was in the transition range between the two signals representing Ep and D. As sensitivity and axial resolution improved, a signal 'dip' or minimum between the Ep and the D signals was recognised, which offered new options on positioning the DEJ \cite{gambichler2006vivo,mogensen2008morphology,blatter2012situ}. In 
2012 the first ultra-high resolution dermatological study was demonstrated, using an early version of the commercial OCT system Skintell, from AGFA \cite{boone2012high,boone2012imaging}. For the first time not only a signal minimum between Ep and D signals but a dark band was displayed in the published images (\cite{boone2012high}, Fig.~4(a) and \cite{boone2012imaging} Fig.~2(b)). Since then, numerous publications based on Skintell was brought forward by Boone et al. studying various skin diseases, but none of these provided detailed analysis on how the different image signals exactly relate to the DEJ \cite{boone2013imaging,boone2013high,boone2014high,boone2015high1,boone2015high2,boone2015high3,boone2016vivo1,boone2016vivo2}. The dark band (DB) analysis in the region between the commonly accepted Ed and D signals, partially presented in Fig.~\ref{UHRcheek} and Fig.~\ref{UHRpalm}, highlights the physical extension of the DB, however more noisy for the palm due to a more attenuated signal further from the skin surface. The DB is observed as a signal floor of the valley, quantified by the double sigmoid fitting. The finite size of the band is quite clear for the UHR-OCT cheek B-scan subsets suggesting it to be up to $28\,\mu m$ thick with an ES less than $6\,\mu m$ for eight subsets out of nine (Fig.~\ref{CheekEval}). The sharp ES suggests that there is a clear transition from one type of skin layer to the other. C-OCT subsets provide five cases (of nine) with non-vanishing DBs of about the same thickness ($5\,\mu m$-$31\,\mu m$) as the UHR-OCT subsets suggest.

For the palm the observed DB for both C-OCT and UHR-OCT has a greater span. Neglecting the three cases (out of 18) with the three largest ESs (B$_3$, C$_3$ and M$_2$ with ESs of $62,\,\mu m$, $32\,\mu m$ and $26\,\mu m$, respectively), also having the three smallest DBs (less than $5\,\mu m$), we observe a consensus amongst the remaining 15 subsets on a DB between $9\,\mu m$ and $76\,\mu m$ of thickness with the ESs of $23\,\mu m$ or less (Fig.~\ref{palmEval}). This means that Also for palm analysis, there is strong indications on a DB located between the Ep and D signals with a thickness detectable for both systems.

With a DB present between the traditionally known Ep and D signals, as seen in Fig.~\ref{UHRcheek}, Fig.~\ref{commcheek}, Fig.~\ref{UHRpalm} and Fig.~\ref{commpalm}, we pose the question; what does the DB signal represent? We find the answer by observing zoom-ins of palm images and recognising the papillaries characteristic for the upper D, which are displayed in Fig.~\ref{pap}. For both systems, but much more blurry for C-OCT, we see the papillary structure as a kind of dark-signal tentacles interfacing the much brighter signal of the stratum spinosum, a part of Ep. With this statement we attribute the DBs found both for the palm and the cheek to the papillary D. This means that the DEJ line in the palm images should be delineated just on the interface between the papillae, seen as a DB, and the stratum spinosum of signal of Ep, as done manually in Fig.~\ref{pap} for HP5 (UHR-OCT). 

Similar for the cheek, the DEJ should be positioned on the upper edge of the DBs. As the UHR-OCT ESs in eight out of nine cases (B-scan subsets), presented in Fig.~\ref{CheekEval}, either are close to or surpassing the resolutions of the system. This suggest that we a dealing with a signal change of sub-resolution nature, which is the case for the stratum basale defining the DEJ.

The DEJ region of normal skin has thoroughly been investigated using RCM as DEJ on these body locations exists less than $100\,\mu m$ under the skin surface and is thus easily accessible for RCM typically exhibiting a penetration depth beyond $200\,\mu m$. The DB detected here is also seen for RCM and supports recognition of the DB as the papillary dermis interfacing with the bright signal of the basal layer and thereby the entrance to Ep \cite{pellacani2007impact,kurugol2015automated}.

\subsection{Dermal papillae in OCT}
The two OCT systems were compared in the setting of recognising different adnexal structures in skin, where the visualisation of the structures are relevant in the context of detecting cancerous tissue. The dermal papillae, already discussed, are seen significantly more clear for the UHR-OCT and in addition one can locate glimpses of intra-papillary structures as marked in Fig.~\ref{pap}. The geometry of the structure is highlighted, when averaging a few neighbouring B-scans with the result seen in Fig.~\ref{Cap}. From these we recognise the small blood vessel structures propagating towards the DEJ inside a dermal papillary and terminating in a curly structure. These are known as capillaries, transporting blood between the arterioles and the venules. Studies with RCM \cite{hegyi2009confocal,archid2012confocal} have observed the capillaries of the dermal papillae of \textit{normal} skin, also as bright structures on a darker signal background. The bright spots on the dark background is provided by blood cells as seen in Fig.~\ref{Cap}. This is in contradiction to capillary sections containing no blood cells and thus appearing as dark lumen-like structures  \cite{cuaruntu2012evaluation,kurugol2015automated}. The latter is seemingly observed in Fig.~\ref{pap}, UHR-OCT, HP5 (marked with a ring) as a bridge between two highly reflecting capillary sections containing blood cells.

In the dermal papillae capillary diameters have been measured in normal skin to be $9.53\pm 1.8\,\mu m$ for RCM (\textit{in vivo}) and $7.5-10\,\mu m$ for electron microscopy \cite{archid2012confocal,braverman2000cutaneous} (\textit{ex vivo}. For the capillaries presented in Fig.~\ref{Cap} we register vessel width dimensions of about 3 pixels ($\sim 9\,\mu m$) laterally and 3-5 pixels ($\sim 4-8\,\mu m$) axially; the difference can be explained by the difference in the lateral and axial resolution. The capillary width dimensions registered are comparable to the previously reported dimensions. With histology as a reference, the capillaries imaged in Fig.~\ref{Cap}, should be compared to the capillaries seen in the histology image of glabrous skin (Fig.~\ref{tyndtykhud}(b)), also presenting capillaries as twisting elongated structures in single papillae reaching towards the skin surface. To our knowledge, no OCT studies have demonstrated \textit{in vivo} imaging of capillaries of \textit{dermal papillae} in human skin before and no \textit{in-vivo} optical imaging studies have demonstrated dermal papillae capillary images in \textit{glabrous} skin before.

An additional example of the difference in image detail due to the difference in resolution is recognised in a \textit{second} DB signal, seen for the palm in the UHR-OCT images between Stratum corneum and stratum spinosum. We denote this DB1 in Fig.~\ref{pap} (marked for HP7). Although we have not analysed DB1 in detail, it is consistent in the UHR-OCT images with similar characteristics of the DB interfacing the DEJ. Again the band nature is harder to recognise in the C-OCT images. 

\subsection{Vellus hair image outcomes of resolution differences}
A third example of the value of a better resolution of an OCT system is that the OCT images would be able to discern the minute vellus hairs in the face, which implies that OCT can detect skin diseases that develop in hair follicles, e.g acne and some skin tumours. We characterised the small hairs in the cheek which we found with high population for HP4. As displayed in Fig.~\ref{hairs}, we know from larger hairs (Fig.~\ref{hairs}(a) and (g)) that the hair itself takes a dark-signal feature relative to the surroundings (being the hair follicle). However smaller hairs, defined here as less than $20\,\mu m$ in width, are invisible for C-OCT but still resolvable for the UHR-OCT. As C-OCT should have the required resolution, which we document for a single hair structure in Fig.~\ref{hairs}(fz), the smallest hair to be detected in the volume, we suspect that the stitching of the multiple images (provided by the multi-focus hardware approach) causes blurring, which is most critical when visualising slanting structures like hairs.
%
\subsection{Shadow compensation and tissue penetration}
As introduced, the UHR-OCT images are shadow compensated and presented on a linear scale where as the C-OCT images are on a logarithmic scale, as common practice, in the attempt to display very weak and very strong signal changes simultaneously. We found shadow compensation to be an important tool, introducing a much better contrast in the image and providing a more realistic signal relations in-depth. An example of the improved contrast when applying shadow compensation is seen in Fig.~\ref{ShadowCompare}, \textit{Appendix C}. The advantage of shadow compensation is very clear when estimating the 8-bit DEJ contrast as done Fig.~\ref{cheek1}(c) and less evident when considering the palm DEJ contrast in Fig.~\ref{palm1}(c), which we associate with the poorer sensitivity roll-off for the UHR-OCT system. 

Another important metric to mention in the system comparison is the penetration depth. We observed generally the C-OCT to provide signals from greater depths, which can be explained by the multi-focus feature. It is however interesting that while the palm images only show little difference in penetration (Fig.~\ref{palm1}), the tissue of the cheek (Fig.~\ref{cheek1}) presents a significant difference in penetration depth in favor of C-OCT. We find this consistent when considering all volumes of cheek and palm, which suggests that a different degree of scattering is indeed taking place in the normal and in the glabrous skin.

\subsection{The significance of the in-depth focal position}
As stated, C-OCT provides $1\,mm$ deep depth of focus comprised by four focii. With a single focusing element the UHR-OCT manages a depth of focus about $0.05\,mm$. For UHR-OCT, positioning of the skin surface at a certain optical path difference was hence critical in order to exploit the full resolution potential at a requested depth in the skin, e.g. in the depth of the DEJ. Positioning of the UHR-OCT focus was only possible with a certainty of a few hundred micrometers in-depth with motion and a non-plane surface complicating the alignment. If motion artefacts were not severe, the effect on the uncertainty in the focus position was minor when globally comparing the image details in the different B-scans. However, zooming in on the $1024\times 1024$ pixel B-scan images, we notice the sensitivity of images details to the focus positioning. An example of this is given for the case of the palm. For HP5, the focus was positioned in the depth of the DEJ and the papillae; this was the only volume providing a high content of intra-papillary structural information of the palm. The remaining HPs on the other hand provided more details and signal either in deeper or in more shallow regions, relative to the dermal papillae. This difference can be noticed in Fig.~\ref{pap} for three HP cases of UHR-OCT where the detail and signal strength from similar anatomy parts vary. We did not observe any difference in the image details when comparing zoom-in B-scans of the C-OCT system.

\section{Conclusion}
We have compared a commercial (C-OCT) and an in-house built system (UHR-OCT) in the frame of healthy skin structures of 11 healthy participants. In combination with shadow compensation we found that UHR-OCT manages to clearly delineate interfaces in the skin due to the improved axial and lateral resolution of $2.2\,\mu m$ and $4.6\,\mu m$, respectively, where similar interfaces are significantly harder to recognise with C-OCT with axial and lateral resolution of $5\,\mu m$ and $7.5\mu m$, respectively. This is surprising given the fact that the C-OCT system uses 4 foci for constant transversal resolution along the depth while for UHR-OCT a simple interface optics was used, with an adjusted focus a few $100\,\mu m$ below the skin surface. The example structures resolved significantly different by the two systems are the dark signal band in the DEJ region, the small hairs of $10-20\,\mu m$ widths, and the dermal papillae. For the latter we have highlighted the advantage of UHR-OCT by imaging capillaries in \textit{glabrous} skin, which to our knowledge has not been achieved before \textit{in vivo} in human tissue by OCT our by any other optical imaging technique.

In more general terms we found UHR-OCT to provide better DEJ contrast for normal skin, also attributed to the applied shadow compensation algorithm whereas C-OCT provided better penetration. 

For the cheek we found DBs in the DEJ regions to exhibit widths of up to about $30\,\mu m$ for both systems where ESs (edge transitions) were seen to surpass the axial resolution of UHR-OCT. 

Imaging DB of the palm showed similar DEJ region contrasts for the two OCT systems, without a noticeable advantage of UHR-OCT. The systems presented strong indications on DBs present and of up to $75\,\mu m$. There was however a very large spread on the DBs which we attribute with poorer signal at the greater depth found for the DEJ in glabrous skin, but also to the coarser axial resolution for the C-OCT system.

By delineating the dark band of the DEJ region, we finally demonstrated visualisation of the exact location of DEJ which we find to be on the interface between the Ep signal and the DB signal. To our knowledge the precise DEJ position in the OCT image has not been reported on despite many publications in dermatology using ultrahigh resolution OCT. With our contribution we hope to add a more precise interpretation of the OCT signals when investigating the details of ultrahigh resolution OCT images.

%
%

\section*{Appendix A: information on voluntary participants}
\begin{table}
\begin{tabular}{|c|c|c|c|}
\hline 
Healthy participant no. & Age & Fitz-Patrick skin type & Sex (female/male) \\ 
\hline 
1 & 33 & 3 & f \\ 
\hline 
2 & 30 & 2 & m \\ 
\hline 
3 & 45 & 3 & f \\ 
\hline 
4 & 27 & 2 & f \\ 
\hline 
5 & 43 & 2 & f \\ 
\hline 
6 & 30 & 2 & m \\ 
\hline 
7 & 72 & 2 & f \\ 
\hline 
8 & 51 & 2 & f \\ 
\hline 
9 & 28 & 2 & m \\ 
\hline 
10 & 28 & 3 & m \\ 
\hline 
11 & 27 & 2 & f \\ 
\hline 
\end{tabular} 
\caption{Overview on age, skin-type and gender information of the healthy participants.}\label{HPtable}
\end{table}

\section*{Appendix B: fitting to the dermo-epidermal region}
\begin{figure}[ht!]
\centering\includegraphics[width=0.8\textwidth]{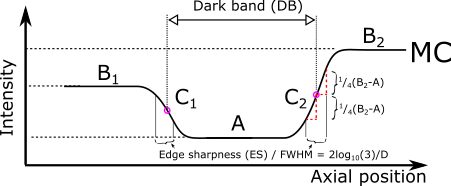}
\caption{Sketch of the double sigmoid model curve (MC) with definitions of the dark band (DB) and the edge sharpness (ES).  $A$, $B_1$, $B_2$, $C_1 $, $C_2$, and $D$ refer to the parameters in eq. \eqref{ESeq}}\label{ModelCurve} 
\end{figure}

In order to extract image measures, a simple model comprised of two sigmoid functions was applied to detect signal levels and signal edge sharpnesses of Ep, DB and D. The model curve $MC$ is expressed as
\begin{align}
MC = A+\frac{(B_1-A)}{1+10^{-\vert D\vert(z-C_1)}}+\frac{(B_2-A)}{1+10^{\vert D\vert(z-C_2)}}\label{ESeq};
\end{align}
The sigmoid function is a step-like function with a smooth transition between two signal levels, the smoothness determined by a transition slope. In our model configuration we have two high-signal plateaus, $B_1$ and $B_2$, and a common low-signal plateau $A$, the latter representing the signal level of DB. The model curve is depicted in Fig.~\ref{ModelCurve}. With the transitions $B_1$ to $A$ and $A$ to $B_2$ we have the associated positions at $C_1$ and $C_2$, respectively, as well as the common steepness of the transition slopes described by $D$. $D$ relates to the  full-width half max (FWHM) of the slope at $C_1$ (or $C_2$) as: FWHM$ = \frac{2}{D}\log_{10}(3)$. In the main text FWHM is referred to as the \textit{edge sharpness} (ES).

\section*{Appendix C: comparison between regular OCT images and images with shadow compensation.}
As the UHR OCT system was built in-house, the image generation and processing was a obvious concern of ours. Like with the C-OCT system, we chose to produce 8-bit images before image analysis was carried out. With 256 signal levels representing the entire 12-bit signal range of the spectrometer, we found that shadow compensation provided better utilisation of the image depth resource than the regular manner of displaying OCT images.
\begin{figure}[h!]
\centering\includegraphics[width=1.0\textwidth]{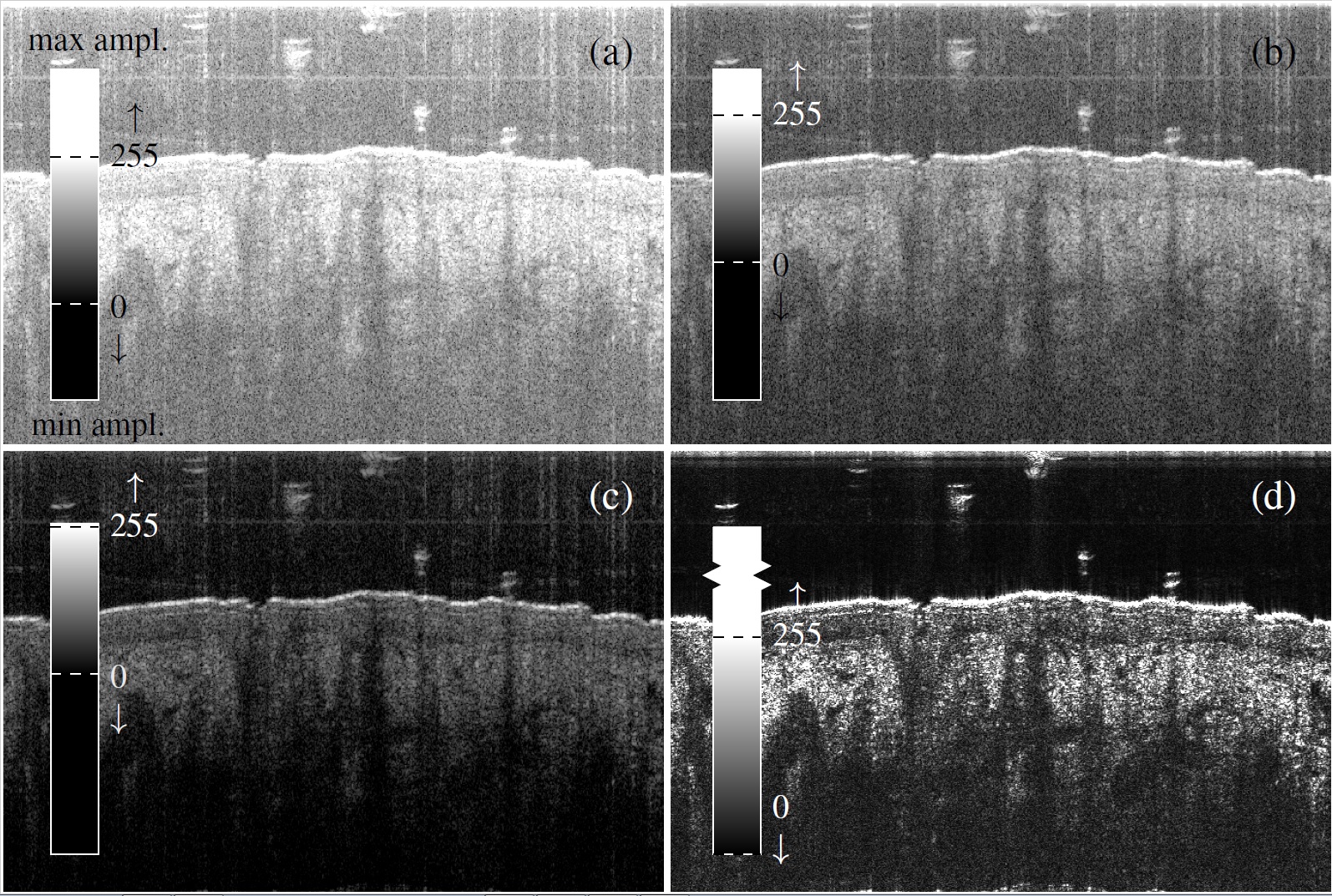}
\caption{Image comparison between commonly logarithmically scaled images and a shadow compensated image. (a)-(c) display the same image using three different 8-bit thresholds on logarithmic scale. (d) shows the same image with shadow compensation and linear scaling.}\label{ShadowCompare}
\end{figure}
In Fig.~\ref{ShadowCompare}(a)-(c), we present three images all on logarithmic scale, but with each there choice of 8-bit thresholding. Fig.~\ref{ShadowCompare}(a) and (c) show saturation in high and low signals, respectively, where (b) represents an intermediate compromise in the attempt to capture both high and low signals. Fig.~\ref{ShadowCompare}(d) show the same image but with shadow compensation which delivers a much better contrast and corrects for exponential signal attenuation associated with traveling through a medium. Please notice that a spectral window (gaussian) was applied for the images seen in Fig.~\ref{ShadowCompare}(a)-(c), but not for (d).

\section*{Appendix D: B-scan averaged data applied for estimating signal contrast of the DEJ region}
\begin{figure}[ht!]
\centering\includegraphics[width=1.0\textwidth]{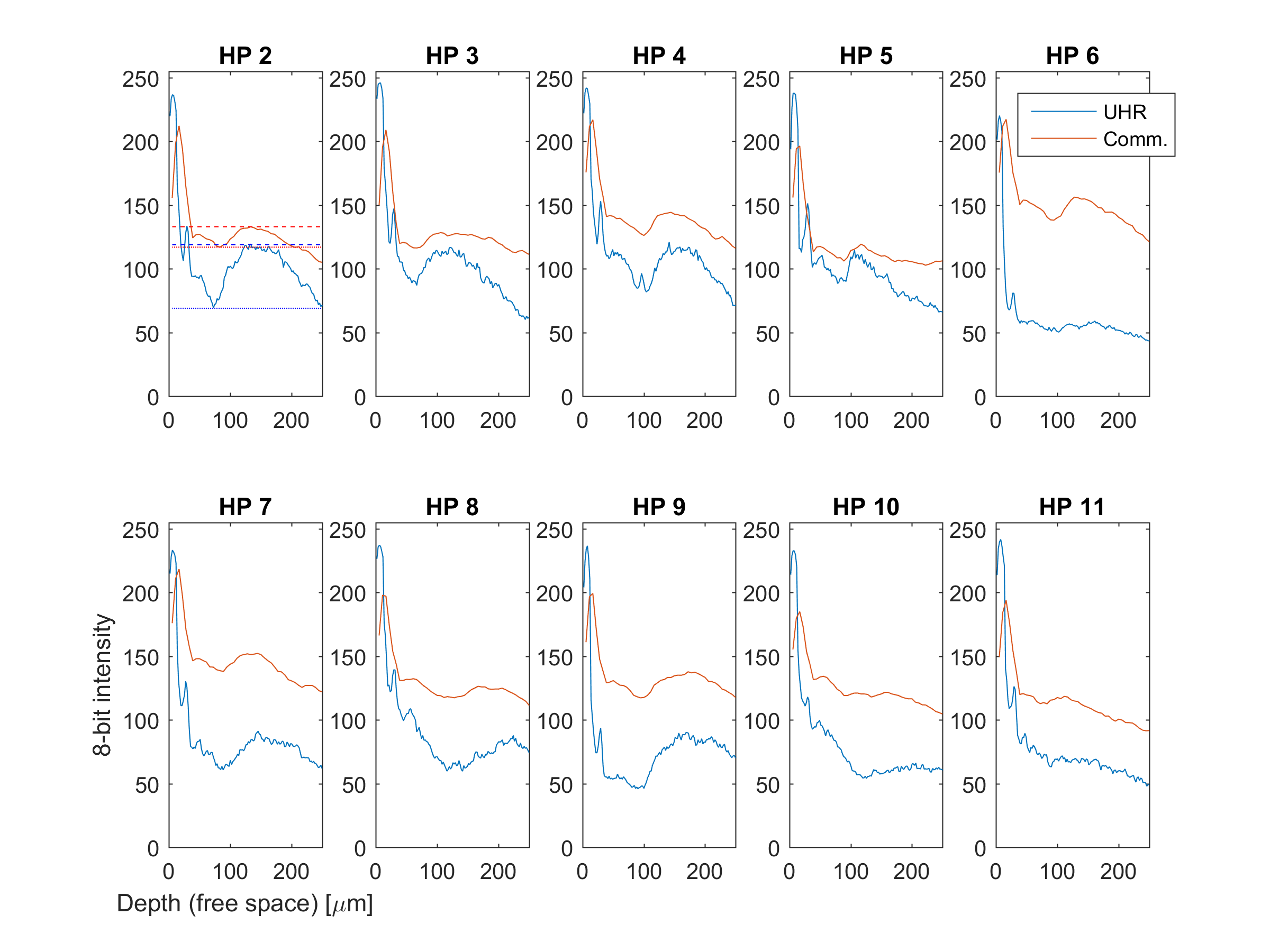}
\caption{Profiles of laterally averaged central B-scans of the cheek for all HPs but HP1 containing severe artefacts. The dashed and solid lines introduced for HP2 mark the respective maximum (D) and minimum (Ep) signal readings applied in determining the DEJ region Ep-D contrast presented in Fig~\ref{cheek1}(c). All depth-axis are optical distances, i.e. scaled as in free space.}\label{HistDataCheek} 
\end{figure}

\begin{figure}[ht!]
\centering\includegraphics[width=1.0\textwidth]{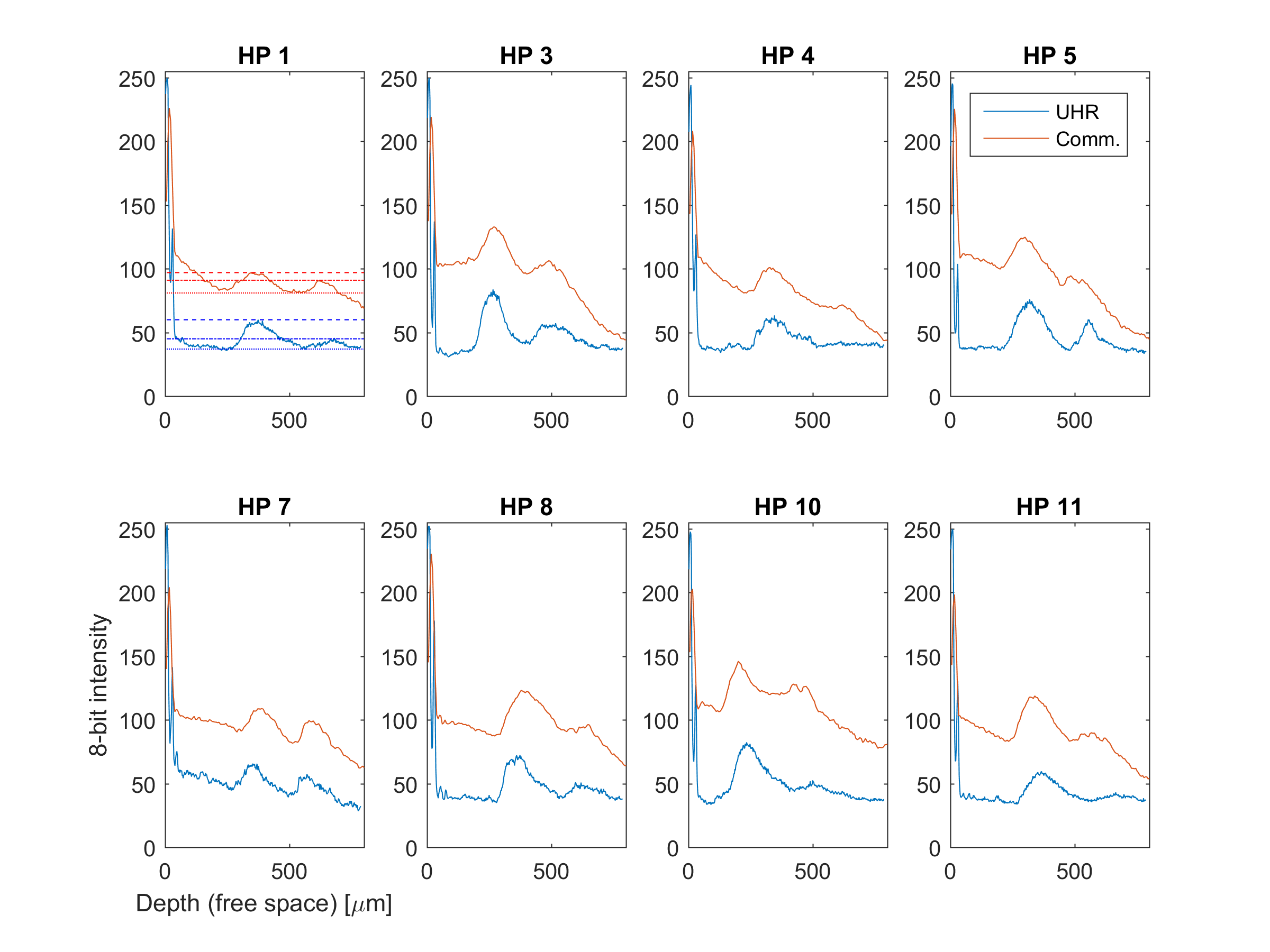}
\caption{Profiles of laterally averaged central B-scans of the palm for all HPs but HP2, HP6 and HP9 containing severe artefacts. The dashed and dotted lines introduced for HP1 mark the respective  Ep and D maxima signal readings where solid lines mark minima (DB) readings. The data is applied in determining the DEJ region Ep-D contrast presented in Fig~\ref{palm1}(c). All depth-axis are optical distances, i.e. scaled as in free space.}\label{HistDataPalm}
\end{figure}

Contrasts of the DEJ region was estimated by averageing the central B-scan of a volume laterally with the surface as reference and detecting maxima and minima commonly associated with Ep and D signals. The histograms presented in Fig.~\ref{cheek1}(c) and Fig.~\ref{palm1}(c) are build from the depth profiles given in Fig.~\ref{HistDataCheek} and Fig.~\ref{HistDataPalm}, respectively. 

\bibliography{References.bib}
\end{document}